\documentclass[12pt,a4paper]{article}
\usepackage{calc}
\usepackage[scale={0.8,0.8},truedimen,centering,paper=a4paper,headsep=10pt,footskip=25pt,includefoot,includehead,heightrounded]{geometry}
\usepackage{titlesec}
\titleformat{\section}{\normalsize\bfseries}{\thesection}{0.5em}{}
\titleformat{\subsection}{\normalsize\bfseries}{\thesubsection}{0.5em}{}
\titlespacing*{\section} {0pt}{2.0ex plus 0.0ex minus 0.0ex}{1.0ex plus 0.0ex}
\titlespacing*{\subsection} {0pt}{1.6ex plus 0ex minus .0ex}{0.8ex plus .0ex}
\usepackage{tocloft}

\usepackage[usenames,dvipsnames]{pstricks}
\usepackage{amsmath,amsfonts,amssymb}
\usepackage[T1]{fontenc}
\usepackage{textcomp}
\usepackage{times}
\usepackage[square,comma,sort&compress,numbers,super,merge,elide]{natbib}
\usepackage{subcaption}
\usepackage[
  pdftitle={Magnetic excitations in some trimeric compounds},%
  pdfauthor={M. Georgiev and H. Chamati},%
  pdfsubject={Statistical Mechanics},%
  pdfkeywords={Molecular magnets},%
  pdfstartview=FitH,%
  bookmarks=true,%
  bookmarksopen=true,%
  breaklinks=true,%
  unicode=true,%
  colorlinks=true,%
  linkcolor=blue,anchorcolor=blue,%
  citecolor=blue,filecolor=blue,%
  menucolor=blue,
  urlcolor=blue]{hyperref}

\usepackage{tikz}			

\definecolor{copper}{rgb}{0.72, 0.45, 0.2}
\definecolor{copper}{RGB}{218,138,103}
\definecolor{copper}{RGB}{184, 115, 51}
\definecolor{copper}{RGB}{153, 102, 102}
	
\begin{document}
	
\title{\normalsize\bfseries MAGNETIC EXCITATIONS IN THE TRIMERIC COMPOUNDS \\
A$_3$Cu$_3$(PO$_4$)$_4$ (A = Ca, Sr, Pb)
}
\author{\normalsize\bfseries
M.~Georgiev and H. Chamati}

\date{}
\maketitle
\begin{abstract} 
We study the magnetic excitations of the trimeric magnetic 
compounds A$_3$Cu$_3$(PO$_4$)$_4$ (A = Ca, Sr, Pb).
The spectra are analyzed in terms of the Heisenberg model and a generic spin
Hamiltonian that accounts for the changes in valence electrons
distribution along the bonds among magnetic ions.
The analytical results obtained in the framework of both Hamiltonians are 
compared to each other and to the available experimental
measurements. The results based on our model show better agreement with the experimental
data than those obtained with the aid of the Heisenberg model.
For all trimers, our analysis reveals the existence of one
thin energy band referring to the flatness of observed excitation peaks.
\end{abstract}

\section{Introduction}

Molecular magnets possess unique 
properties and are ideal candidates for exploring 
the interplay of the quantum and the classical worlds.
They may manifest a great variety of magnetic features
determined from weakly interacting isolated
fundamental structural units, such as dimers, trimers and tetramers
\cite{furrer_magnetic_2013}.
The effect of quantum tunneling in single-molecule magnets
\cite{wernsdorfer_exchange-biased_2002,schenker_phonon_2005} and the
response of spin-switching in the frustrated antiferromagnetic
chromium trimmer \cite{jamneala_kondo_2001} are some prominent examples.
With their short-range spin correlation the small spin clusters stand as 
elegant tools for studying the relevant coupling processes. 
Magnetic measurements on trimer copper chains A$_3$Cu$_3$(PO$_4$)$_4$ 
with (A = Ca, Sr), reported in Ref. \cite{drillon_1d_1993},
show that the intertrimer exchange couplings are negligible and thus the
trimers might be considered as separate clusters.
These results were confirmed
via INS experiments \cite{matsuda_magnetic_2005,podlesnyak_magnetic_2007}
that shed light on the magnetic spectra with the aid of the antiferromagnetic 
Heisenberg model involving nearest and next-nearest intratrimer interactions, 
and later they were extended to the compound Ca$_3$Cu$_3$(PO$_4$)$_4$
\cite{furrer_magnetic_2013}. Moreover, it turns out that the interaction 
between edged spins in isolated trimers is also negligible.
The difference in the magnetic properties among the compounds 
Ca$_3$Cu$_2$Ni(PO$_4$)$_4$ \cite{ghosh_spin_2012} and
Ca$_3$Cu$_2$Mg(PO$_4$)$_4$ \cite{ghosh_magnetic_2010} is another 
demonstration for the richness of the physical features arising from a symmetrically
trivial linear spin trimers, see e.g. Ref. \cite{ghosh_nmr_2010}.

In the present article we report a theoretical study of the magnetic spectra 
of magnetic clusters. We focus our attention on the trimeric compounds
A$_3$Cu$_3$(PO$_4$)$_4$ with (A = Ca, Sr, Pb), for which the magnetic
excitations are determined experimentally \cite{matsuda_magnetic_2005,podlesnyak_magnetic_2007}.
To describe the magnetism in the compounds A$_3$Cu$_3$(PO$_4$)$_4$ we
employ the approach devised in Refs.
\cite{georgiev_mexchange_2019,georgiev_epjb_2019}.
The approach is based on a generic spin
Hamiltonian that allows to compute effectively the changes of electron's 
density distribution along the complex exchange bridges among magnetic
centers. We compare the results of our study obtained in the
framework of the named generic spin Hamiltonian and its Heisenberg 
counterpart demonstrating their equivalence and differences.

The rest of this paper is structured as follows: In Section
\ref{sec:fundamprinc} we present the keystone relations for the neutron scattering
intensities and formulate explicitly the generic spin Hamiltonian.
In Sections \ref{sec:copper}
we explore the low-lying magnetic excitations of the
compounds A$_3$Cu$_3$(PO$_4$)$_4$ (where A stands for Ca, Sr, Pb)
within the framework of the Heisenberg model and our Hamiltonian.
A summary of the results obtained throughout this paper along with
conclusions are presented in Section \ref{sec:conclusion}.

\section{The model and the method}\label{sec:fundamprinc}
\subsection{Inelastic Neutron Scattering}

To determine
the energy level structure and the transitions corresponding to
the experimentally observed magnetic spectra one needs a 
number of parameters to account for all couplings in the system. 
It is cumbersome
to apply a general approach with a unique set of parameters
that can describe all possible magnetic effects and in addition to
distinguish between inter-molecular and intra-molecular features. 
Usually, one starts with bilinear spin microscopic models, such as the Heisenberg 
Hamiltonian \cite{furrer_magnetic_2010} and depending on the exhibited 
magnetic features different interaction terms are included \cite{chamati_theory_2013}.

To obtain meaningful results one calculates
the neutron scattering intensities $I_{n'n}(\mathbf{q})$ integrated over 
the angles of scattering vector $\mathbf{q}$ of the neutron.
For identical magnetic ions, represented by the operators
$\hat{s}^\alpha_i$ and $\hat{s}^\alpha_j$, they read
\cite{lovesey_1986,malcolm_neutron_1989,furrer_neutron_2009,toperverg_neutron_2015}
\begin{equation}\label{eq:ScateringIntensities}
I_{n'n}(\mathbf{q}) \propto F^2(\mathbf{q}) \sum_{\alpha, \beta} \varTheta^{\alpha \beta}
S^{\alpha \beta} (\mathbf{q},\omega_{n'n}),
\end{equation}
where $F(\mathbf{q})$ is the spin magnetic form factor \cite{jensen_rare_1991},
$\varTheta^{\alpha \beta}$ is the polarization factor
and $\alpha, \beta, \gamma \in \{x,y,z\}$.
In \eqref{eq:ScateringIntensities} the magnetic
scattering functions are explicitly written as
\begin{equation}\label{eq:ScatterFunctions}
S^{\alpha \beta} (\mathbf{q},\omega_{n'n}) = 
\sum_{n,n',i,j}^{} e^{\mathrm{i} \mathbf{q} \cdot \mathbf{r}_{ij}} 
p_n \langle n \lvert \hat{s}^\alpha_i \rvert n' \rangle \langle n'\rvert 
\hat{s}^\beta_j \rvert n \rangle
\delta (\hbar \omega_{n'n} - E_{n'n}), 
\end{equation}
where $\omega_{n'n}$ is the frequency of a magnetic excitation related
to a transition between the states $\rvert n \rangle$ and $\rvert n' \rangle$ 
with the corresponding energy $E_n$ and $E_{n'}$, respectively.
Further, $e^{\mathrm{i} \mathbf{q} \cdot \mathbf{r}_{ij}}$ is the structure factor 
associated with the cluster geometry, $p_n = Z^{-1} e^{-E_n/k_B T}$ is the 
population factor (with $Z$ the partition function).

\subsection{The generic spin model} \label{sec:spinmodel}
 
The distribution 
of coupled magnetic centers (ions) plays a crucial
role in uniquely determining the scattering intensities.
Even when these effective bonds are indistinguishable with respect to
their lengths and the total spin, according to
\eqref{eq:ScatterFunctions}, one can obtain different in 
magnitude neutron scattering intensities. However, to identify each
intensity one has to use an appropriate
spin model leading to an energy sequence such that the 
$\delta$ function in the r.h.s of \eqref{eq:ScatterFunctions} 
defines the relevant spin bonds 
with respect to the structure factors.

To describe the magnetic spectra in the considered trimeric compounds 
we employ the proposed in Ref. \cite{georgiev_mexchange_2019,georgiev_epjb_2019}
generic spin Hamiltonian
\begin{equation}\label{eq:AddHamiltonian}
\hat{\mathcal{H}} = 
\sum\limits_{i \ne j}^{} J_{ij} 
\hat{\boldsymbol{\sigma}}_i \cdot \hat{\mathbf{s}}_j,
\end{equation}
where the couplings $J_{ij} = J_{ji}$
are effective exchange constants and the operator
$\hat{\boldsymbol{\sigma}}_i \equiv
(\hat{\sigma}^x_i, \hat{\sigma}^y_i, \hat{\sigma}^z_i)$ 
accounts for the differences in valence electron's distribution
with respect to the $i$th magnetic center. Let us note that model
\eqref{eq:AddHamiltonian} was applied successfully to explore the magnetic spectra of 
the molecular magnet Ni$_4$Mo$_{12}$ \cite{georgiev_epjb_2019}.

\section{Magnetic spectra of the trimers A$_3$Cu$_3$(PO$_4$)$_4$ (A =Ca, Sr and Pb)} 
\label{sec:copper}
\subsection{The Hamiltonian}
\begin{figure}[ht!]
	\centering
	\includegraphics[scale=1]{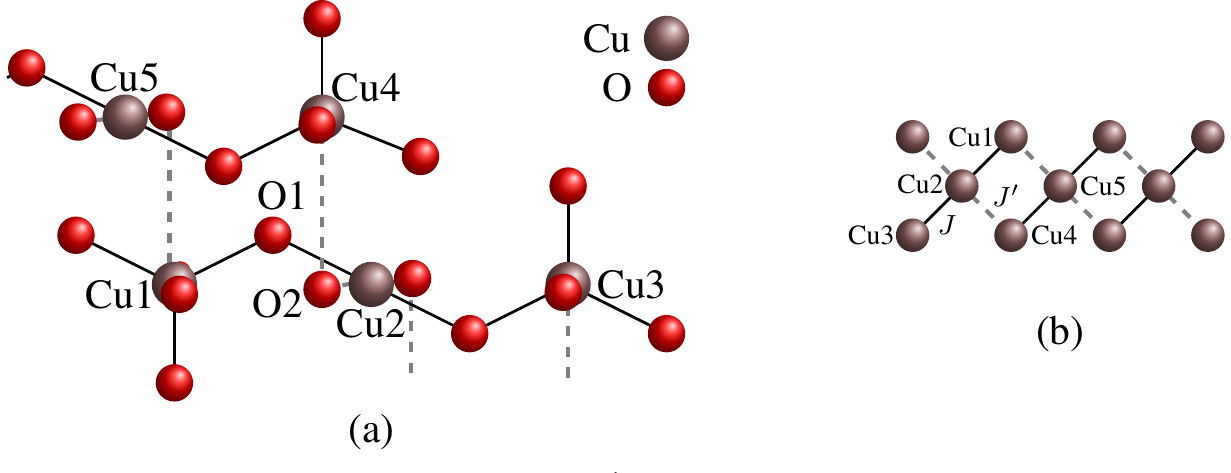}
	\caption{(a)
		Exchange pathways in A$_3$Cu$_3$(PO$_4$)$_4$ (A = Ca, Sr, Pb).
		Copper colored circles represent copper ions,
		the red ones stand for oxygen atoms. The solid (black) and dashed
		(gray) lines represent the intratrimer and intertrimer exchange pathways,
		respectively. (b) Schematic representation of the intratrimer $J$ and
		intertrimer $J'$ magnetic interactions in the array of isolated trimers.
		\label{fig:Copper}}
\end{figure}

The magnetic compounds A$_3$Cu$_3$(PO$_4$)$_4$ (A = Ca, Sr, Pb) are
convenient spin trimer systems for testing the Hamiltonian 
\eqref{eq:AddHamiltonian}.
On Fig. \ref{fig:Copper} (a) we show a small fragment of the
copper ions structure with the relevant exchange pathways with respect
to the arrangement of oxygen atoms. Whence the Cu2 ion is surrounded by four oxygen atoms
on a plane, while Cu1 and Cu3 ions are surrounded by five oxygen atoms
constructing a distorted square pyramid. For the sake of clarity the other elements
are not shown and only two oxygen atoms along the intratrimer
Cu1--O1--Cu2 and intertrimer Cu2--O2--Cu4 pathways are labeled. In
general, the exchange processes appear to be more complex and
depend on the global structure of the compounds \cite{drillon_1d_1993}. 
Besides the superexchange interactions are sensitive \cite{matsuda_magnetic_2005} to 
the angle between Cu$^{2+}$ bonds and their lengths suggesting that the 
intertrimer Cu2--Cu4 interaction is much smaller than the
intratrimer ones, \textit{i.e.} Cu1--Cu2 and Cu3--Cu2. Thus, the intertrimer
exchange can be neglected and the Cu$^{2+}$ sub-lattice is
considered as a one-dimensional array of isolated
spin trimers Fig. \ref{fig:Copper} (b).

%
%

Taking into account that Cu1-Cu2 and Cu2-Cu3 are bonded by a single 
oxygen ion we set $J_{ij}\to J_{12}=J$ and perform a study of the magnetic 
excitations.
Owing to the trimer symmetry we apply the coupling scheme 
$|s_2-s_{13}|\le s \le |s_2+s_{13}|$, where $s$ and $s_{13}$ 
(with $|s_1-s_3|\le s_{13} \le |s_1+s_3|$) are the trimer 
and Cu1-Cu3 coupled pair spin quantum numbers, respectively.
Thus, the Hamiltonian \eqref{eq:AddHamiltonian} reads
\begin{equation}\label{eq:CopperHamilton}
\hat{\mathcal{H}} = J \left( \hat{\boldsymbol{\sigma}}_{13}\cdot\hat{\mathbf{s}}_2 + 
\hat{\boldsymbol{\sigma}}_2\cdot\hat{\mathbf{s}}_{13} + 
\hat{\boldsymbol{\sigma}}_1\cdot\hat{\mathbf{s}}_3+
\hat{\boldsymbol{\sigma}}_3\cdot\hat{\mathbf{s}}_1\right). 
\end{equation}
With respect to the selected spin coupling scheme the
total spin eigenstates are denoted by $\rvert s_{13},s,m\rangle$.

\subsection{Energy levels}

%
%
\begin{figure*}[b!]
\centering
\includegraphics[scale=1]{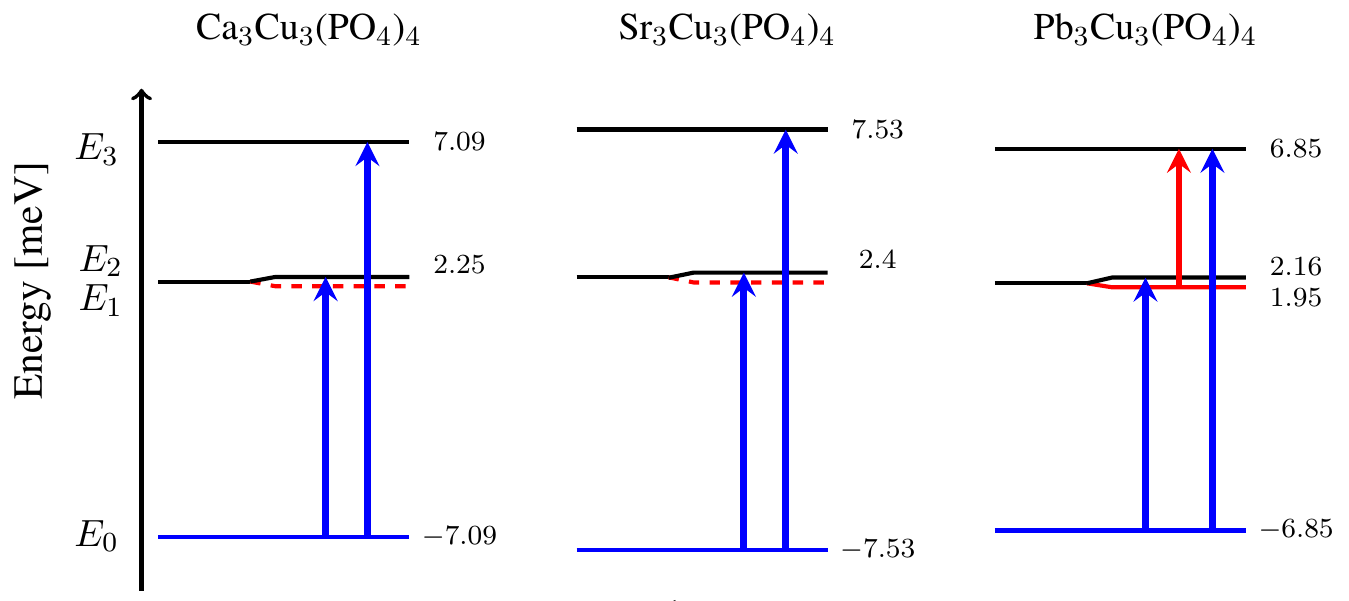}
\caption{Energy level structure of the compounds
A$_3$Cu$_3$(PO$_4$)$_4$ (A = Ca, Sr, Pb). The blue arrows
show the ground state transitions, while the red arrow stands for the excited
transition. The energy levels corresponding to the ground state are
designated by blue lines.
The initial energy level of the excited transition is depicted by
a red line, while by analogy to Pb$_3$Cu$_3$(PO$_4$)$_4$ the dashed red lines
stand for a presumed second sub level of the excited doublet level.
}
\label{fig:CopperEnergyStructure}
\end{figure*}

The isolated trimer is described by four quartet and four
doublet eigenstates. The eigenvalues of \eqref{eq:CopperHamilton} are
denoted by $E^m_{s_{13},s}$. Further, analyzing the energy spectrum we obtain 
the ground state energy for $s_{13}=1$, $s=\tfrac12$. 
The respective doublet states are
$\big\lvert 1, {\tfrac{1}{2}}, \pm\tfrac{1}{2} \big\rangle$
with corresponding energies
\begin{equation}\label{eq:CopperEnergyGS}
E^{\pm 1/2}_{1,1/2} = -\tfrac32 J.
\end{equation}
The second pair of doublet states is
associated with the first excited energy level, see Fig.
\ref{fig:CopperEnergyStructure}. 
The edged spins of the isolated trimer are coupled in a singlet,
with corresponding state 
$\big\lvert 0, {\tfrac{1}{2}}, \pm\tfrac{1}{2} \big\rangle$.
Now, using
\eqref{eq:CopperHamilton} we end up with
\begin{equation}\label{eq:CopperEnergyFirstEx}
E^{\pm 1/2}_{0,1/2} = - \tfrac32J a^{0,0}_{13},
\end{equation}
where the parameter $a^{0,0}_{13}\in\mathbb{R}$ account for the 
variations of electrons spatial distribution along the Cu1-Cu3 
exchange bridge.
To fully characterize the experimentally observed transitions 
for Pb$_3$Cu$_3$(PO$_4$)$_4$ one requires at least three excited 
energy levels. Bearing in mind that the quartet level is
four-fold degenerate, we
deduce that the corresponding coefficient may take only two values 
$a^{0,0}_{13}\in\left\lbrace c^1_{13},c^2_{13}\right\rbrace$. 
Further, the observed excitations spectra
\cite{matsuda_magnetic_2005} are not broadened signaling that
$\big|c^1_{13}-c^2_{13}\big| \approx 0$.
Therefore, taking into account \eqref{eq:CopperEnergyFirstEx} we get
\[
E^{\pm 1/2}_{0,1/2} = -\tfrac32 J c^n_{13} \quad n=1,2.
\] 	
For all four quartet eigenstates 
$\big\lvert 1, {\tfrac{3}{2}}, m \big\rangle$, with
$m=\pm\tfrac12, \pm\tfrac32$, we have
\[
E^{\pm 1/2}_{1,3/2}=E^{\pm 3/2}_{1,3/2} = \tfrac32 J.
\]
The energy sequence consists of four levels. 
Henceforth we denote these levels as follow 
\begin{equation}\label{eq:CopperEnergyNotation}
E_0 = -\tfrac32 J, \quad E_1 = -\tfrac32 J c^1_{13},
\quad E_2 = -\tfrac32 J c^2_{13}, \quad E_3 = \tfrac32 J.
\end{equation}

\begin{figure}[!h]
\centering
\includegraphics[scale=1]{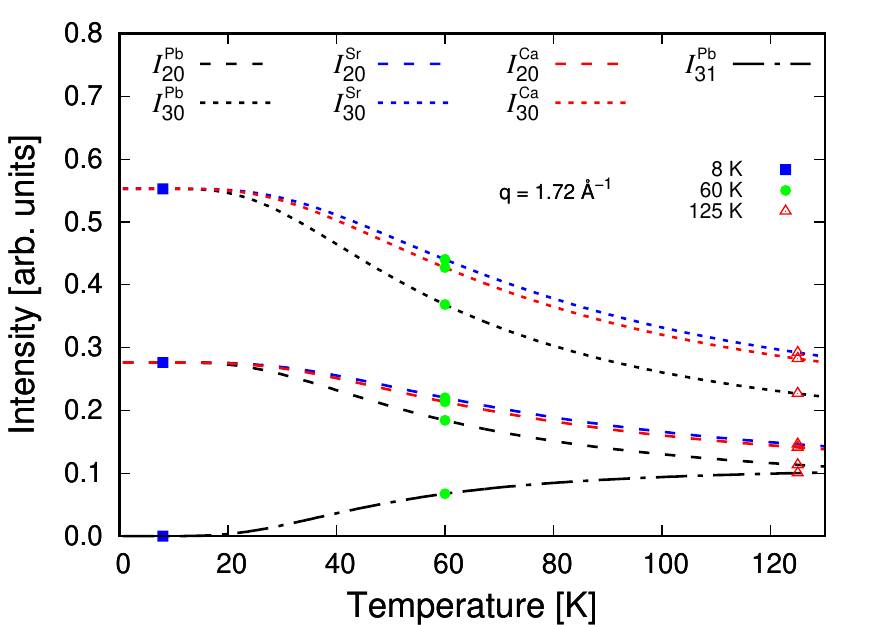}
\caption{
Scattering intensities $I^{\mathrm{A}}_{20}$, $I^{\mathrm{A}}_{30}$ and
$I^{\mathrm{Pb}}_{31}$
with (A = Ca, Sr, Pb) as a function of the temperature, calculated with
the Hamiltonian \eqref{eq:AddHamiltonian}. 
The blue squares, the green
circles and red triangles correspond to the values of the intensities
given in Tab. \ref{tab:CopperIntensThreeTemp}.
\label{fig:CuIntensity}}
\end{figure}

Therefore, we have at hand the parameters $J$ and $c^{n}_{13}$.
The coupling $J$ accounts for the interaction along Cu1-Cu2 and 
Cu2-Cu3 bridges and $c^{n}_{13}$ will indicate any changes in 
the interaction between edged ions. However, we take further 
actions and derive the following relation
$J_{c^{n}_{13}}=J\big(\tfrac{3}{4}c^{n}_{13}+\tfrac{1}{4}\big)$,
where $J_{c^{n}_{13}}$ represents the exchange constant between 
the next-nearest neighbors.

\subsection{Scattering intensities}	
Based on the selection rules
$\Delta s_{13} = 0, \pm 1$, $\Delta s = 0, \pm 1$ and $\Delta m = 0,
\pm 1$ and the aid of the identities 
$S^{\alpha\beta}(\mathbf{q},\omega_{n'n})+S^{\beta
\alpha}(\mathbf{q},\omega_{n'n})=0$, $S^{\alpha
\alpha}(\mathbf{q},\omega_{n'n})=S^{\beta
\beta}(\mathbf{q},\omega_{n'n})$, where $\alpha\ne\beta$ and $n,n' = 0,1,2,3$, we may
compute the scattering functions.
Moreover, taking into account the cluster structure,
we have $\sum_{\alpha}\Theta^{\alpha \alpha} = 2$.
The analysis of intensities reported in \cite{matsuda_magnetic_2005} 
allows us to determine the observed first magnetic excitation.
It corresponds to the transition between the ground state energy $E_0$
and $E_2$ with scattering functions
\[
S^{\alpha \alpha}(\mathbf{q},\omega_{20}) =\tfrac{1}{3} 
[1-\cos(2\mathbf{q}\cdot\mathbf{r})]p_0,
\]
where $\mathbf{r}$ is the vector of the average distance $r$ between
neighboring ions with $\mathbf{r}_{31}=2\mathbf{r}$. 
The degeneracy of the quartet energy level is four--fold 
and hence the second ground state excitation refers to transition 
from the doublet 
$\big\lvert 1, {\tfrac{1}{2}}, \pm\tfrac{1}{2} \big\rangle$
to the quartet states $\big\lvert 1, {\tfrac{3}{2}}, m \big\rangle$,
where $m=\pm\tfrac12, \pm\tfrac32$. 
Hence, for $E_0\to E_3$ we get
\[
S^{\alpha \alpha}(\mathbf{q},\omega_{30}) = \tfrac{2}{9} 
[3 + \cos(2\mathbf{q}\cdot\mathbf{r}) - 4\cos(\mathbf{q}\cdot\mathbf{r})]p_0.
\]
The excited peak is indicated by the transition $E_1\to E_3$.
The corresponding scattering functions are
\[
S^{\alpha \alpha}(\mathbf{q},\omega_{31}) = \tfrac{2}{3}[1 - 
\cos(2\mathbf{q}\cdot\mathbf{r})]p_1.
\]
Therefore, according to \eqref{eq:ScateringIntensities}
we estimate the relevant intensities obtaining
\begin{equation}\label{eq:CopperIntegIntensities}
\begin{array}{ll}
I_{20} \propto  \gamma_{20}
 \left[ 1-\frac{\sin(2qr)}{2qr}\right] 
 F^2(q),
&
I_{30} \propto \gamma_{30} 
 \left[ 1 + \frac{\sin(2qr)}{6qr} - 4\frac{\sin(qr)}{3qr}\right] 
 F^2(q),
\\[0.3cm]
I_{31} \propto  \gamma_{31} 
 \left[ 1-\frac{\sin(2qr)}{2qr}\right]
  F^2(q),
  &{}
\end{array}
\end{equation}
where
$\gamma_{20} = \tfrac23 p_0$, $\gamma_{30} = \tfrac{12}9 p_0$
and  $\gamma_{31} = \tfrac43 p_1$.
Moreover, for dications Cu$^{2+}$ the form factor reads
$F(q) = 256/(16+q^2 r^2_{\mathrm{o}})^2$, 
where $q$ is the magnitude of the scattering vector, 
$r_{\mathrm{o}} = 0.529 \, \mathrm{\mathring{A}}$ is the Bohr radius.

\begin{figure}[h!]
\begin{subfigure}{\linewidth}
\centering
\includegraphics[scale=1]{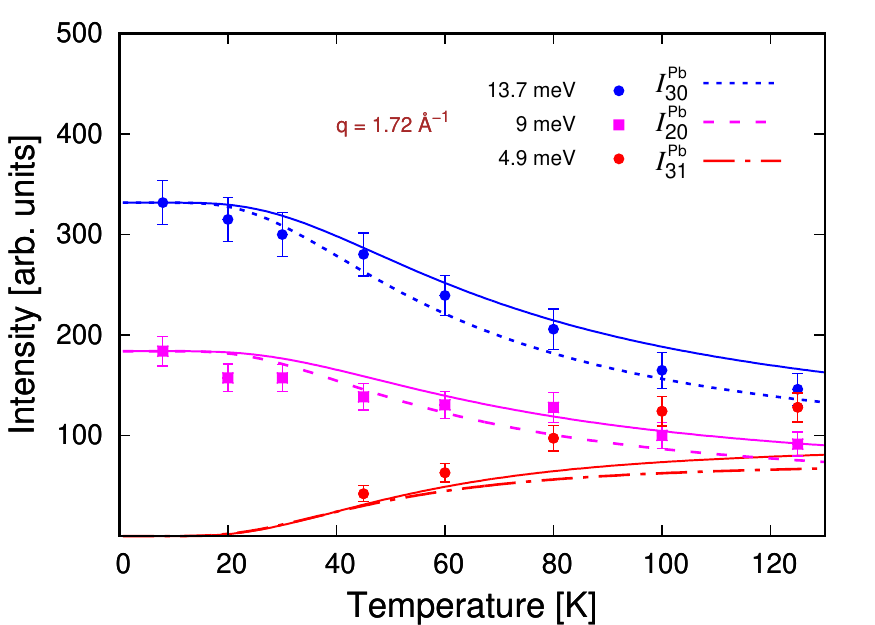}
\subcaption{Scattering intensities for the compound 
Pb$_3$Cu$_3$(PO$_4$)$_4$ as a function of the temperature,
along with experimental results from Ref. \cite{matsuda_magnetic_2005}.
The solid and dashed lines show the calculated intensities
for the Heisenberg model and Hamiltonian
\eqref{eq:AddHamiltonian}, respectively.
\label{fig:PbIntensity}}
\end{subfigure}
%
\begin{subfigure}{\linewidth}
\centering
\includegraphics[scale=1]{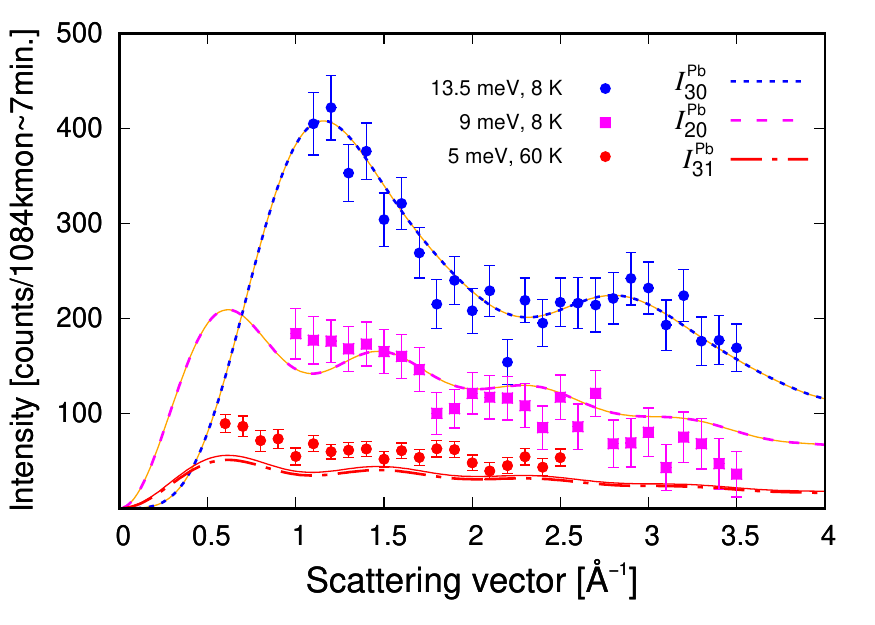}
\subcaption{
Calculated intensities as a function of the scattering vector
for Pb$_3$Cu$_3$(PO$_4$)$_4$ along with the experimental
data of Ref. \cite{matsuda_magnetic_2005}. 
The dashed lines depict the intensities obtained from the 
Hamiltonian \eqref{eq:AddHamiltonian}. 
The solid red and orange lines correspond to the Heisenberg model.
$I^{\mathrm{Pb}}_{20}$ and $I^{\mathrm{Pb}}_{30}$ correspond to the ground state
transitions at $T=8$ K.
The intensity $I^{\mathrm{Pb}}_{31}$ stands for the excited transition
at $T=60$ K.
\label{fig:CuFFIntensity}
}
\end{subfigure}
\caption{Scattering intensities. \label{fig4}}
\end{figure}

\subsection{Energy of the magnetic transitions}	
Taking into account \eqref{eq:CopperEnergyNotation} and 
\eqref{eq:CopperIntegIntensities} for the transition energies we get
\begin{equation}\label{eq:CopperGSTransitions}
E_{20} = \tfrac32 J \left(1 - c^2_{13}\right) ,
\quad
E_{30} = 3 J,
\quad
E_{31} = \tfrac32 J \left( 1 + c^1_{13}\right) .
\end{equation}
Neutron
scattering experiments performed on Pb$_3$Cu$_3$(PO$_4$)$_4$ with
$T \geq 60$ K \cite{matsuda_magnetic_2005} show the
presence of a third peak at about $4.9$ meV, which may be related
to the excited transition energy $E_{31}$. 
The values of $c^1_{13}$,
$c^2_{13}$ and $J$, according to INS experiments
\cite{matsuda_magnetic_2005} performed on polycrystalline samples
A$_3$Cu$_3$(PO$_4$)$_4$ (A = Ca, Sr, Pb) are shown in Tab.
\ref{tab:Copper}. In addition, for the compound Ca$_3$Cu$_3$(PO$_4$)$_4$
we have $c^2_{13} = -0.32(8)$ and $J \approx 4.741 \, \mathrm{meV}$
based on INS data at $T=1.5$ K
\cite{furrer_magnetic_2010,podlesnyak_magnetic_2007}.
\begin{table}[ht!]
\caption{
The values of the coupling constants and the quantities $c^1_{13}$,
$c^2_{13}$ for our model applied to
A$_3$Cu$_3$(PO$_4$)$_4$ (A = Ca, Sr, Pb)
obtained by taking into account the experimental data 
in Ref. \cite{matsuda_magnetic_2005}. 
\label{tab:Copper}
}
\begin{center}
\begin{tabular}{ccccccccc}
\hline
\hline
 A & $E_{20}$ & $E_{30}$  
& $E_{31}$ & $c^1_{13}$ & $c^2_{13}$ & $J$ & $J_{c^2_{13}}$ & $J_{c^1_{13}}$ \\
\hline
Ca & 9.335 & 14.174 & $-$ & $-$ & -0.317 & 4.725 & 0.058 & --\\
Sr & 9.936 & 15.064 & $-$ & $-$ & -0.319 & 5.021 & 0.054 & --\\
Pb & 9.005 & 13.693 & 4.9 & -0.284 & -0.315 & 4.564 & 0.062 & 0.168\\ 
\hline
\hline
\end{tabular}
\end{center}
\end{table}

The temperature dependence of the integrated scattering intensities for each
compound is shown on Fig. \ref{fig:CuIntensity}.  On
Fig. \ref{fig:PbIntensity} we present the scattering intensities for 
Pb$_3$Cu$_3$(PO$_4$)$_4$ computed with our Hamiltonian and
the Heisenberg model along with the
experimental data taken from Ref. \cite{matsuda_magnetic_2005}. 
Let us point out that
our results are in better agreement with their
experimental counterpart for $I_{20}^\mathrm{Pb}$ and $I_{30}^\mathrm{Pb}$, 
while for $I_{31}^\mathrm{Pb}$
we have a qualitative agreement. The averaged
magnitudes of the scattering vector $q$ and the distance $r$ between
neighboring ions are taken from Ref.
\cite{matsuda_magnetic_2005}, $q=1.72\ \mathrm{\mathring{A}}^{-1}$
and $r = 3.6 \ \mathrm{\mathring{A}}$. The explicit expressions of the
scattering intensities for each transition are
\begin{equation}
\label{eq:CrossSectionsCopper}
I^{\mathrm{A}}_{20}(T) \propto 0.5528 Z^{-1}_\mathrm{A} e^{-\frac{E^{\mathrm{A}}_0}{k_BT}},
\quad
I^{\mathrm{A}}_{30}(T) \propto 1.1057 Z^{-1}_\mathrm{A} e^{-\frac{E^{\mathrm{A}}_0}{k_BT}},
\quad
I^{\mathrm{Pb}}_{31}(T) \propto 1.1056 Z^{-1}_{\mathrm{Pb}} e^{-\frac{E^{\mathrm{Pb}}_1}{k_BT}},
\end{equation}
where A = Ca, Sr, Pb. As $T$ vanishes the scattering intensities
$I^{\mathrm{A}}_{20}$ and $I^{\mathrm{A}}_{30}$ are equal by about a
factor of 2, see Tab. \ref{tab:CopperIntensThreeTemp}.
\begin{table}[!ht]
\caption{
Calculated values of integrated scattering intensities $I^{\mathrm{A}}_{n'n}$
[arb. units], using the Hamiltonian \eqref{eq:AddHamiltonian}, for the trimers
A$_3$Cu$_3$(PO$_4$)$_4$ (A = Ca, Sr, Pb) at temperatures 8, 60
and 125 K, depicted on Fig. \ref{fig:CuIntensity}.
\label{tab:CopperIntensThreeTemp}
}
\begin{center}
\begin{tabular}{lccc}
\hline
\hline
$T$ [K]                &    8     &     60   &    125   \\
\hline
$I^{\mathrm{Ca}}_{20}$ & 0.276(4) & 0.213(7) & 0.141(2) \\
$I^{\mathrm{Ca}}_{30}$ & 0.552(8) & 0.427(4) & 0.282(5) \\
$I^{\mathrm{Sr}}_{20}$ & 0.276(4) & 0.220(2) & 0.146(1) \\
$I^{\mathrm{Sr}}_{30}$ & 0.552(8) & 0.440(5) & 0.292(2) \\
$I^{\mathrm{Pb}}_{20}$ & 0.276(4) & 0.184(3) & 0.113(4) \\
$I^{\mathrm{Pb}}_{30}$ & 0.552(8) & 0.368(6) & 0.226(8) \\
$I^{\mathrm{Pb}}_{31}$ & 0        & 0.067(3) & 0.100(3) \\	
\hline
\hline
\end{tabular}
\end{center}
\end{table}	
For $T > 20$ K a third peak
sets in, but the evaluated intensity $I^{\mathrm{Pb}}_{31}$ remains
smaller than the experimentally observed one
\cite{matsuda_magnetic_2005}. In contrast to the functions
$I^{\mathrm{Pb}}_{30}$ and $I^{\mathrm{Pb}}_{20}$ the intensities of
the ground state transitions for A = Ca, Sr
decrease slowly with temperature. The predicted peak for
Pb$_3$Cu$_3$(PO$_4$)$_4$ is in concert with
the experimental findings \cite{matsuda_magnetic_2005}. 
Unfortunately there are no experimental data confirming the presence of this third peak for
the compounds Ca$_3$Cu$_3$(PO$_4$)$_4$ and Sr$_3$Cu$_3$(PO$_4$)$_4$
and hence the energy level $E_1$ could not be included in determining the sequence of energy spectrum.
On Fig. \ref{fig:CopperEnergyStructure} the presumed energy levels 
$E^{\mathrm{Ca}}_1$ and $E^{\mathrm{Sr}}_1$ are illustrated with dashed red lines.
For all compounds the scattering intensities as a function of the
magnitude of the scattering vector are represented in Fig. \ref{fig:CuFFIntensity}.

\section{Conclusion}\label{sec:conclusion}
We propose an study for the magnetic excitations of the compounds 
A$_3$Cu$_3$(PO$_4$)$_4$ with (A = Ca, Sr, Pb).
To this end, we use a generic bilinear spin Hamiltonian \eqref{eq:AddHamiltonian}
that accounts for the variations in the electron's spatial distributions
along the exchange bridges. 
Alongside with the named Hamiltonian we compute the magnetic spectrum
in the framework of the Heisenberg
Hamiltonian and compare the outcome from both models, 
see Figs. \ref{fig:PbIntensity} and \ref{fig:CuFFIntensity}. We found
that the results obtained with our model are in better agreement with
the INS experimental data
\cite{matsuda_magnetic_2005,podlesnyak_magnetic_2007} 
than the Heisenberg model. On the other hand
our results for the Heisenberg model coincide with those reported by
other authors 
\cite{matsuda_magnetic_2005,podlesnyak_magnetic_2007,furrer_magnetic_2010}.

With respect to the energy levels sequence and relevant eigenstates
the Heisenberg and our Hamiltonian \eqref{eq:AddHamiltonian} lead to 
similar values.
For the investigated compounds, the ground state energy is related to the Cu1-Cu3 triplet bond, and
the neutron energy loss, associated to both ground state magnetic 
excitations, is due to the local triplet-singlet transition.
However, the spin Hamiltonian
\eqref{eq:CopperHamilton}, with the intrinsic parameter
$a^{0,0}_{13}$,
identifies the experimentally observed third peak (about 4.9 meV) for
the compound Pb$_3$Cu$_3$(PO$_4$)$_4$
\cite{matsuda_magnetic_2005} accurately, while 
the Heisenberg model is enable to reproduce it.
We obtain one thin energy band composed of two very close 
energy levels that corresponds
to the Cu1-Cu3 singlet state (see e.g. Fig. \ref{fig:CopperEnergyStructure}).
The energy band width signals for the small change in the electrons distribution 
along the Cu1-Cu2-Cu3 bridge due to the temperature.
Thus, the intensities indicated
by dashed lines on Fig. \ref{fig:PbIntensity} decrease rapidly than in
the case of the Heisenberg model.
In other words, the inequality
$\big|c^n_{13}\big|<1$ for $n=1,2$, shows that in the doublet level, 
the spatial distribution of the electrons common to
the edge ions is such that the exchange becomes negligible. 
Further, it points out that the next-nearest neighbor coupling 
slightly varies with respect to the temperature taking two values
$J_{13}\in\big\{J_{c^1_{13}},J_{c^2_{13}}\big\}$, see Tab. \ref{tab:Copper}.
On the other hand the difference $\big|c^1_{13}-c^2_{13}\big| = 0.031$ explains the
sharpness of the experimentally observed peaks 
\cite{matsuda_magnetic_2005,podlesnyak_magnetic_2007}.

\subsection*{Acknowledgments}
The authors are indebted to Prof. N. Ivanov and
Prof. J. Schnack for very helpful discussions, and 
to Prof. M. Matsuda for providing us with the experimental data used
in FIGs. \ref{fig:PbIntensity} and \ref{fig:CuFFIntensity}.
This work was supported by the Bulgarian National Science Fund under
contract DN/08/18.


%

\vspace*{0.8cm}

\parindent 0cm

Institute of Solid State Physics, \\
Bulgarian Academy of Sciences, \\
Tsarigradsko Chauss\'ee 72, \\
1784 Sofia, Bulgaria \\
mgeorgiev@issp.bas.bg

\end{document}